\documentclass[review]{elsarticle}

\usepackage{lineno,hyperref} 
\pdfoutput=1 
\modulolinenumbers[5]
\usepackage{multirow,multicol} 
\usepackage{soul} 
\usepackage{amssymb} 
\usepackage{amsthm}
\usepackage{amsmath}

\usepackage{natbib}
\usepackage{url} 
\usepackage[T1]{fontenc} 

\usepackage{geometry} 

\usepackage{tabularx}
\usepackage{adjustbox}
\usepackage{longtable}
\usepackage[table]{xcolor} 

\usepackage{setspace}

\usepackage{graphicx} 

\usepackage{booktabs}

\usepackage[utf8]{inputenc} 

\usepackage{enumitem} 
\newcommand{\bulletlabel}{\raisebox{0.2ex}{\small$\bullet$}}
\newlist{enumtable}{enumerate}{6}
\setlist[enumtable]{label=\bulletlabel, topsep=0.0in, leftmargin=0.1in, rightmargin=0.1in, itemsep= 0in, partopsep=0in}

\usepackage{nomencl}
\makenomenclature

 \usepackage{float}
\usepackage{babel}

\usepackage{enumitem}
\usepackage{chngpage}
\makenomenclature

\usepackage{array}
\newcolumntype{P}[1]{>{\centering\arraybackslash}p{#1}}
\newcolumntype{M}[1]{>{\centering\arraybackslash}m{#1}}
\usepackage{mathtools}

\usepackage{array,amsfonts}

\begin{document}

\begin{frontmatter}

\title{Unsupversied feature correlation model to predict breast abnormal variation maps in longitudinal mammograms}

\author[1]{Jun Bai}
\author[2]{Annie Jin}
\author[2]{Madison Adams}

\author[2,3]{Clifford Yang}

\author[1]{Sheida Nabavi}
\cortext[cor1]{Corresponding author: 
Sheida Nabavi, sheida.nabavi@uconn.edu;
  Tel.: +1-860-486-0756}

\address[1]{Department of Computer Science and Engineering, University of Connecticut, 371 Fairfield Way, Storrs, CT 06269, USA}
\address[2]{University of Connecticut School of Medicine, 263 Farmington Ave. Farmington, CT 06030, USA}
\address[3]{Department of Radiology, UConn Health, 263 Farmington Ave. Farmington, CT 06030, USA}

\begin{abstract}

Breast cancer continues to be a significant cause of mortality among women globally. Timely identification and precise diagnosis of breast abnormalities are critical for enhancing patient prognosis. In this study, we focus on improving the early detection and accurate diagnosis of breast abnormalities, which is crucial for improving patient outcomes and reducing the mortality rate of breast cancer. To address the limitations of traditional screening methods, a novel unsupervised feature correlation network was developed to predict maps indicating breast abnormal variations using longitudinal 2D mammograms. 
 
The proposed model utilizes the reconstruction process of current year and prior year mammograms to extract tissue from different areas and analyze the differences between them to identify abnormal variations that may indicate the presence of cancer. The model is equipped with a feature correlation module, an attention suppression gate, and a breast abnormality detection module that work together to improve the accuracy of the prediction. The proposed model not only provides breast abnormal variation maps, but also distinguishes between normal and cancer mammograms, making it more advanced compared to the state-of-the-art baseline models. The results of the study show that the proposed model outperforms the baseline models in terms of Accuracy, Sensitivity,  Specificity, Dice score, and cancer detection rate.

\end{abstract}


\end{frontmatter}

\printnomenclature
\hfill

\section{Introduction}
\label{sec:introduction}

Breast cancer is one of the top health concerns for women in the world \citep{siegel_cancer_2020}. The high chance of developing breast cancer has also led to considerable anxiety among the general public. Research has demonstrated that early detection is crucial for reducing the burden of breast cancer. Healthcare professionals have emphasized the importance of early detection to improve the survival rate \citep{coleman_early_2017}. In particular, mammography is widely recognized as the gold standard for breast cancer screening for asymptomatic women to identify breast cancer before any symptom appears \cite{tabar_beyond_2001,duffy_relative_2003}. American Cancer Society (ACS) recommending annual mammography screening for women ages 40 and older since 2004 \citep{smith_american_2003}. Although annual screening of mammograms increases breast cancer survival rate, the large volume of screening mammograms and the use of double readings put a strain on the healthcare system and can compromise its efficiency. Furthermore, the complexity of mammograms can pose challenges for radiologists, potentially resulting in false and missed diagnoses \citep{huynh_false-negative_1998,murphy2007analysis}. Besides, small tumors surrounded by dense fibroglandular breast tissue may go undetected, which can delay diagnosis and prevent early detection \citep{huynh_false-negative_1998}. Moreover, false positive results are possible, which can lead to unnecessary procedures such as benign biopsies, additional expenses, and psychological distress for patients.
\citep{ong_national_2015,nelson_harms_2016}.          

In clinical practice, the traditional computer-aided detection (CAD) system --- relies on image processing and pattern recognition techniques --- provides a second opinion on breast cancer detection and image interpretation to radiologists. The CAD system involves handcrafted features ---using conventional image processing methods--- from mammograms and then classifies them using statistical methods. In contrast, deep learning, a subset of machine learning, has been increasingly used over the past decade to develop CAD systems that can automatically extract features from raw inputs using multiple layers of neural networks. In recent years, many studies have utilized deep learning, especially convolutional neural networks (CNNs), for mammogram classification\citep{teare_malignancy_2017,hang_glimpsenet_nodate,lotter_multi-scale_2017,ribli_detecting_2018,sahiner_classification_1996,shih-chung_b_lo_multiple_2002,kooi_comparison_2016,t_discriminating_2017,abbas_deepcad_2016,huynh_digital_2016,wu_deep}. In many deep learning mammogram studies, Full-Field Digital Mammography (FFDM) has been utilized to learn the data distribution, although some studies have used reduced-size images \citep{teare_malignancy_2017,hang_glimpsenet_nodate,lotter_multi-scale_2017,ribli_detecting_2018,wu_deep}, patches, or regions of interest (ROI) to minimize computational complexity \cite{sahiner_classification_1996,shih-chung_b_lo_multiple_2002,kooi_comparison_2016,t_discriminating_2017,abbas_deepcad_2016,huynh_digital_2016}. Although, deep learning based models have shown a remarkable ability to carry out a wide range of medical imaging tasks including breast cancer screening \cite{teare_malignancy_2017,hang_glimpsenet_nodate,lotter_multi-scale_2017,ribli_detecting_2018,sahiner_classification_1996,shih-chung_b_lo_multiple_2002,kooi_comparison_2016,t_discriminating_2017,abbas_deepcad_2016,huynh_digital_2016,wu_deep} and have made automated breast cancer mammography screening feasible, developing such a model still remains challenging \cite{bai2021applying}, especially for localizing abnormalities in mammograms due to the unique characteristics of mammogram images. 

Breast abnormalities are small and share features with the background tissue, making it difficult to accurately detect tumors with ambiguous boundaries or different shapes. 
 Radiologists commonly use previous mammogram scans as a reference point to make an inference when examining current mammogram scans in order to spot suspicious tissue. Not only in clinical practice but also in studies it has been shown that comparing current mammogram scans with prior ones improves the outcomes of mammography classification. In the study by Kooi and Karssemeijer twin CNNs are used to classify mammogram ROIs by comparing primary mammograms with their corresponding history mammograms or opposite breast mammograms. However, the model requires some level of radiologist knowledge to identify the mass ROIs and to map them from current mammograms to their prior ones \citep{kooi_classifying_2017}. In another study by Perek et al., CNN models were used to extract features from current and previous year images, and long short-term memory (LSTM) layers were employed to learn the differences and classify mammograms. However, this method requires image registration during data pre-processing, and training LSTM with a small dataset to learn tissue changes can cause overfitting \citep{perek_learning_2019}. In our previous work, we developed a feature fusion siamese network to predict cancer incorporate with patients' prior and current image\citep{bai2022feature}. While some studies have suggested using prior mammograms to classify breast cancer, no studies have been proposed to utilize prior knowledge to locate breast cancer.

In addition to the challenge of extracting effective features from complex breast tissue, the analysis of mammograms is constrained by a small sample size, particularly with accurately annotated, datasets \citep{zhou_review_2021, bai2021applying}. 
Mammogram data labeling presents a number of formidable challenges, chief among which is the requirement for specialized domain knowledge to ensure accurate and reliable labeling. Compounding this challenge is the considerable variability in mammogram quality, resolution, and orientation, rendering the accurate pixel-level identification and labeling of regions of interest a complex and arduous task. This directly leads to heavy human labor for data annotation and non-generalized annotation. One solution to the challenge of the non-annotated data and the difficulty of extracting useful features from complex breast tissue is to use an unsupervised learning approach. The breast cancer screening technique, which involves comparing current mammogram images with prior ones, naturally lends itself to building an unsupervised learning model that can learn the differences between images through the image reconstruction process. In this study, in light of U-Net's~\citep{long2015fully,siddique2021u}  empirical effectiveness and the reality of mammogram data scarcity, we propose UFCN, an \underline{u}nsupervised \underline{f}eature \underline{c}orrelation \underline{n}etwork, to predict a map showing  abnormal variations by comparing the history of mammograms, which we call it  abnormal variation map (AVM). Specifically, inspired by the radiologist's reading procedure, we propose a parallel feature correlation encoder to compares high resolution prior and current mammograms to  predict and visualize tumor location without using annotated training data. The contributions of this paper are summarized as follows:

\begin{itemize}
    \item We introduce a novel model that can detect breast cancer without requiring pixel level labeled data and that covers the entire process from start to finish.
    \item We present a novel feature correlation module that can accurately pinpoint feature discrepancies between the current and previous image.
    \item We propose a novel attention suppression gate to enhance the model's capacity in differentiating between normal and cancer cases.
    \item We propose a novel module to predict abnormal maps of the breast, which can help localizing tumors.
    \item To the best of our knowledge, this is the first time an unsupervised model has been used to predict a breast abnormality variation map by utilizing patients' current and prior mammograms.  
\end{itemize}

\section{Related Study}
Breast cancer localization using FFDM has been extensively studied using deep learning algorithms. Researchers have explored various advanced models to achieve accurate localization of breast cancer, including convolutional neural networks (CNNs), recurrent neural networks (RNNs), and hybrid models. Despite the success of supervised deep learning approaches, unsupervised deep learning algorithms have limited implementation in breast cancer localization. This section will review studies that employed deep learning to localize breast cancer and unsupervised segmentation studies. 

\subsection{Breast Cancer localization studies}

In the case of breast cancer localization, the goal is to determine the precise location of the abnormal object within the cancerous patient's image. The main model --- end-to-end model --- used to detect cancer is CNN \citep{platania2017automated,sun2016preliminary,akselrod2016region,al2017detection}. Several researchers have introduced different approaches to detect and locate different types of abnormalities in mammography images such as region-based network (R-CNN) for ADs \citep{ertosun2015probabilistic}, dual CNN-based visual search system for mass \citep{kisilev2016medical}, multi-scale deep belief network \citep{carneiroautomated}, self-transfer learning model \citep{hwang2016self}, U-Net for all types of cancer\citep{abdelhafiz2020convolutional} and adversarial model for mass \citep{zhu2016adversarial}. These methods involve pre-processing, image enhancement, and normalization techniques to improve detection and localization accuracy. 

In many mammogram localization studies, patches are often used to overcome computation challenges \citep{sun2016preliminary,dhungel2015deep,choukroun2017mammogram}. In the study by \citeauthor{choukroun2017mammogram}, the authors proposed detecting microcalcifications (Cals) using discriminative local information contained in patches. The patch-based approach has some limitations, as the input patches are obtained from non-overlapping regions, leading to potential difficulties in precisely localizing masses. Furthermore, the small size of the input patches can create challenges in distinguishing between normal and abnormal tissue, which can negatively impact the accuracy of the detection and diagnosis of medical conditions. These issues highlight the need for continuing research and the development of more effective approaches for breast cancer detection. 

Besides the aforementioned data-driven models, there are a few general models to localize breast cancer such as the YOLO-based models \citep{platania2017automated,al2017detection}, UNet-based models \citep{baccouche2021connected}, and attention UNet-based models \citep{li2019attention}. The remarkable success achieved by these cutting-edge models in localizing cancer cannot be overstated. However, even the most advanced models have a few limitations that must be addressed. One particularly significant limitation is the difficulty in accurately predicting results for normal patients. This is because these models are designed to operate under the assumption that patients have already been diagnosed with cancer or are suspected of having cancer. As a result, the models are trained using only cancer and benign data, making their prediction ability for normal patients uncertain. In addition, the supervised training process for these models presents a challenge. Annotating breast tumors is a costly and unclear process due to the varying and non-standardized tumor location annotation process Different levels of experience and practice in breast cancer diagnostics could cause different annotation styles. For example, unlike other object annotations, Cals --- a type of breast cancer screened as clusters of tiny bright dots ---poses a challenge in determining whether to annotate the entire area or each individual dot.

\subsection{Unsupervised segmentation studies}

Unsupervised segmentation is a rapidly emerging field that has attracted the attention of researchers across various disciplines, including medical image analysis. Through meticulous investigation, researchers have developed cutting-edge techniques to cluster similar pixels and segment objects in natural images by assigning pixel-level labels to images \citep{kanezaki2018unsupervised,kim2020unsupervised}. Despite this, the application of these methods in medical image analysis poses significant challenges as the pixel intensity fails to exhibit significant differences, rendering this approach ineffective. Unlike objects in nature images, cancer and disease present a multitude of distinct characteristics during their development stages, presenting an opportunity to apply innovative unsupervised segmentation techniques that could revolutionize the field of medical image analysis.

In the world of clinical diagnostics, it is essential to assess a patient's disease development to predict any possible occurrence of a particular ailment. In other words, by screening patients' current condition, a synthetic history screen can be generated to predict any abnormal variation areas. In a pioneering study by \citeauthor{van2022unsupervised}, the authors employed a deep generative model to generate prior images using patients' lesion-contained current-year images \citep{van2022unsupervised}. This approach involved using the difference map as segmentation for abnormal lesions. Furthermore, this technique has also been applied in breast cancer localization, as outlined in the research by \citeauthor{park2023unsupervised}. In their work, the authors proposed using StyleGAN2 to generate patients' synthesized normal images based on their current images, which aids in localizing abnormal tissue. The researchers trained StyleGAN2 on normal mammograms to predict abnormal tissue on the difference between generated synthetic normal images and their corresponding suspicious current images\citep{park2023unsupervised}. Pioneering models have demonstrated the feasibility of unsupervised learning in disease localization. However, there are certain limitations when it comes to generating prior images to learn the difference map as a disease segmentation technique. One such limitation is that normal prediction remains unknown, which poses a significant challenge to researchers.  

\section{Methods}
\label{sec:Method}

The proposed model in this study, which we call it  Unsupervised Feature Correlation Network (UFCN), takes advantage of the deep U-shaped residual connected autoencoder reconstruction process to learn the abnormal variation maps. In the following section, we present our model method in detail.

\subsection{Definition}

In this study, we define inputs as
$I_i = \{(C_i \in \mathbb{R}^2, P_i\in \mathbb{R}^2\}$, where $C_i$ is a current year image that can be biopsy-confirmed cancer or normal and $P_i$ is a prior image that is normal. 
Let $f : \mathbb{R}^2 \rightarrow \mathbb{R}^p$ be a feature extraction function and $g : \mathbb{R}^p \rightarrow \mathbb{R}^2$ be a feature transpose function. For an $f$ and $g$, the estimate of ${C}_i$, $\hat{C}_i$, can be  obtained using these functions.
 On the other hand, if  $f$ and $g$ can be learned through training for a given  $\hat{C}_i$ then learning $f$ and $g$ can be considered as a typical image reconstruction problem. 
If functions $f$ and $g$ have the property of being differentiable, then their parameters can be optimized using gradient descent.
However, our study aims to predict the abnormal variation map (AVM) by training $f$ and $g$ in an unsupervised manner. Particularly, the proposed model will be trained to learn $f$ and $g$ under the task of reconstruction and prediction of the AVM using the learned feature difference between $C_i$ and $P_i$. To put this into practice, we solved two sub-problems:  training  parameters of $f$ and $g$ with  given $\hat{C}_i$ and prediction of the optimal AVM with learned $f$ and $g$. Because we also want to predict no abnormal changes for  normal patients ($C_i$s) where AVM should detect nothing,  we introduced a mapping function $h:\mathbb{R}^p \rightarrow \mathbb{R}$ in $g$ to map the probability  of binary labels of $C_i$s, $y_i$s. In our case, the $y_i$ represents binary label of either normal or cancer of current mammograms. 
 
\subsection{Unsupervised feature correlation network (UFCN)}

\begin{figure*}
\centering
   \includegraphics[width=1.1\textwidth]{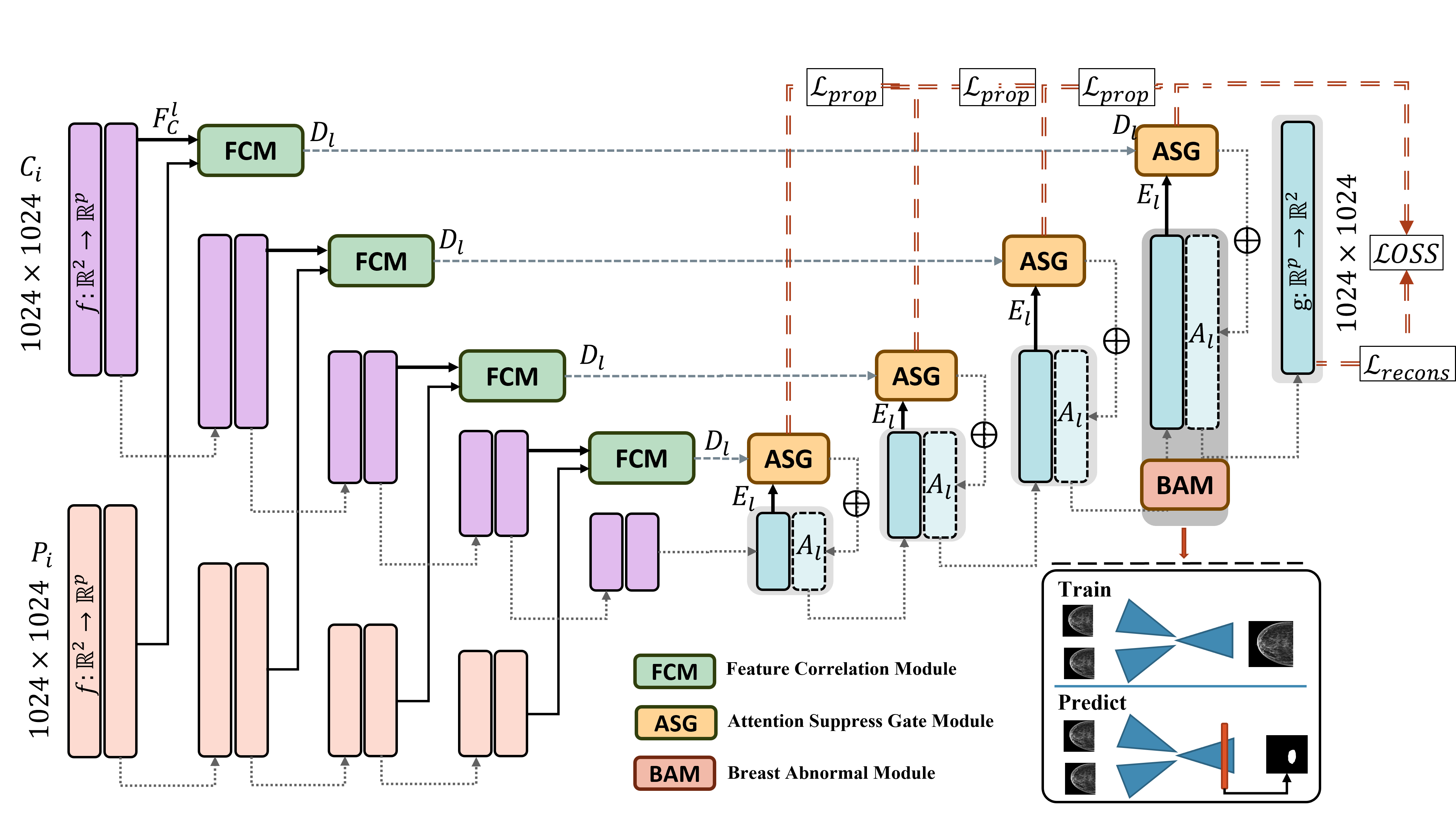}
    \caption{Overview of the proposed model's structure. The model consists of two encoders with inputs $C_i$ (current mammogram) and $P_i$ (previous mammogram), and one decoder output $\hat{C}_i$. The green panel is the feature correlation map module (FCM), which computes the signal difference between features extracted using function  $f$ from current and previous mammograms at each layer. The yellow panel is attention suppress gate (ASG). The red panel is the breast abnormality detection map module (BAM), which predicts the abnormal variation maps (AVM) as output. $\oplus$ is concatenation, and $\ominus$ is matrix subtraction.}
    \label{fig:model}
\end{figure*}
Our proposed model,  UFCN, is an unsupervised CNN-based model. UFCN consists of an identical parallel twin encoder and a reconstruction decoder (Figure~\ref{fig:model}). UFCN embeds feature correlation modules (FCMs)  into encoder layers and attention suppress gate modules (ASGs) into decoder layers. The breast abnormality map detection module (BAM) is embedded at the decoder's layer $L-1$, where $L$ is the total number of layers, to generate AVMs. A pair of current and prior images are input into the two identical parallel CNN blocks to learn the features. In each corresponding CNN block, the learned features of pairs of images using function $f$ are fed into FCMs to learn  differences between current and prior images and their features. The output of FCM, $D_l$,  represents the breast tissue changes from prior mammograms ($P_i$) to current mammograms ($C_i$). $D_l$s  residual connect to ASG at the decoder stage. The output of the decoder is the reconstruction of current mammograms, $\hat{C}_i$.

Selecting an activation function to trigger the model neurons in an unsupervised learning model  is critical. Hence, we employed three activation functions in this study: 1) ReLU activation function ($\sigma_1$) shown in equation ~\ref{equ:RELU}, sigmoid activation function ($\sigma_2$) shown in equation ~\ref{equ:sig}, and SiLU activation function ($\sigma_3$) shown in equation ~\ref{equ:silu} \citep{elfwing2018sigmoid}. 

\begin{equation}
\label{equ:RELU}
    \sigma_1(x) = max(0, x),
\end{equation}
\begin{equation}
\label{equ:sig}
    \sigma_2(x) = \frac{1}{1 + e^{-x}},
\end{equation}
 \begin{equation}
 \label{equ:silu}
     \sigma_3(x) = x\cdot \sigma_2(\beta x),
 \end{equation}
where $\beta$ is a trainable parameter and $x$ is input feature. The SiLU activation function has a desirable characteristic known as self-stabilization. This means that the point at which the derivative of the function is zero acts as a ``soft floor'' for the weights, which helps to regulate the learning process by discouraging the development of excessively large weights. We implemented the SiLU activation in the FCM modules and CNN blocks in the encoder and decoder; the ReLU activation in the ASG modules for faster gradient descent; and the sigmoid activation in the ASG and BAM modules. 

\subsection{Feature correlation module (FCM).}

To take advantage of the paired images through the reconstruction process, we embedded FCM modules into each layer of the encoder stage. FCM (Figure~\ref{fig:module}.a.) learns multi-scaled feature correlation between current and prior mammograms to learn newly grown tumors. The FCM module output, $D_l \in \mathbb{R}^p$, can be expressed as $D_l = \sigma_3(F_C^l) \ominus \sigma_3(F_P^l)$, where $F_C^l = f(C_i) \in \mathbb{R}^p$ and $F_P^l = f(P_i)\in \mathbb{R}^p$ are the feature maps of the current and prior mammograms at layer $l$,  $\ominus$ is matrix subtraction, and $\sigma_3$ is the SiLU activation function \citep{elfwing2018sigmoid}.    

\subsection{Attention suppress gate module (ASG).}

\begin{figure*}
\centering
   \includegraphics[width=1.\textwidth]{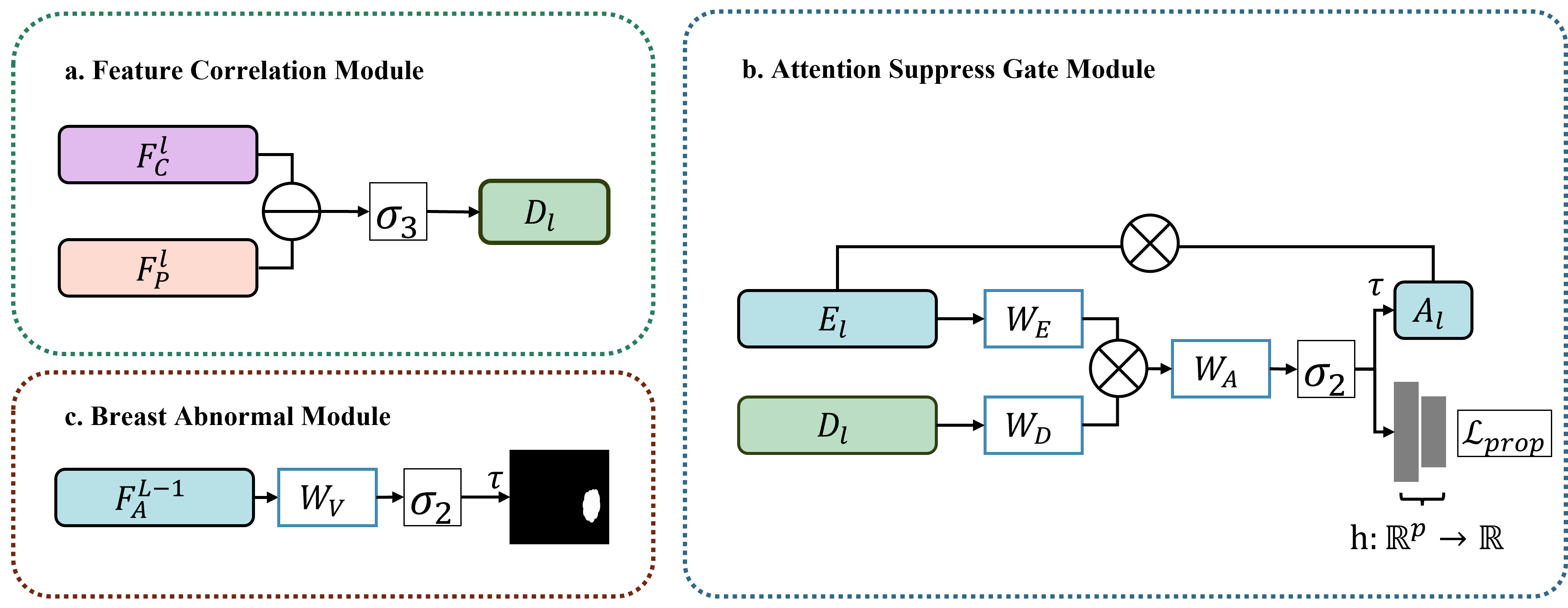}
    \caption{a. Block diagram of the feature correlation module (FCM). $F_C^l$ and $F_P^l$ are current year and previous year feature maps, respectively, generated in the layer $l$ of the decoder  (function $f$). $D_l$ is output from FCM in layer $l$. The $\ominus$ is matrix subtraction. b. Block diagram of the attention suppresses gate module (ASG). $E_l$ is feature generated in encoder  (function $g$) layer $l$, $D_l$ is output of FCM, and $A_l$ is attention coefficients. The $\otimes$ is Hadamard product. c. Block diagram of the breast abnormality detection module (BAM). $F_A^l$ is the feature map from layer $L-1$.}
    \label{fig:module}
\end{figure*}

The ASG module is embedded into each layer of the decoder stage at the image $C_i$ reconstruction process with function $g$. The ASG (Figure~\ref{fig:module}.b.) aims to selectively highlight specific areas of an image through attention weightings. In our specific problem, we also utilize ASGs to completely remove any activation from the normal patients' images. As the model is trained, the soft attention aspect will prioritize regions with greater weights while the hard suppresser --- mapping function $h$ --- eliminates any activity from the normal images.

ASG adds weight to breast tissue areas and reduces the impact of changes in breast borders. ASG at layer $l$ outputs attention coefficients $A_l \in \mathbb{R}^p$ whiles takes $D_l$ and $E_{l-1} \in \mathbb{R}^p$ as input, where $E_{l-1}$ is decoder output at  layer $l-1$. $A_l$ is computed as $A_l = \sigma_2(\sigma_1(W_E^T E_{l-1} \otimes W_D^T D_l) W_A^T)W_E^TE_{l-1}$, where $\sigma_2$ is the sigmoid activation function,  and $\otimes$ is Hadamard product. To have more aggressive soft attention, we introduced a threshold $\lambda$ to suppress the attention coefficients lower than the threshold and to remain the attention region for the attention coefficients higher than the threshold. To prevent neurons from dying, we suppress the lease-activated region (the feature map region below the threshold) to a small constant value instead of zero.    

The hard suppressor acts as a regularizer that maps region-activated feature map to probability of $\hat{y}$ as $\hat{y} = h(A_l) = \sigma_2(W_f^T A_{l}) \in \mathbb{R}$ where $W_f \in \mathbb{R}^p$ is a vector of trainable parameters in our study. The mapped probability of normal and cancer,$\hat{y}$, are participated in loss term to compute the gradient. ASG progressively suppresses features responding in irrelevant background regions and normal patient's image. The output of ASG at each layer $l$ is concatenated with its corresponding encoder features at each layer $l$, and then the features are fed forward to the next layer until reaching the last layer that reconstructs $C_i$ and outputs $\hat{C_i}$.

\subsection{Breast abnormality detection module (BAM).}

As we defined in the previous section, the optimal AVM is predicted with learned $f$ and $g$. To achieve an accurate binary mask AVM indicating the abnormal regions, we embedded  BAM in the decoder stage. The BAM module (Figure~\ref{fig:module}.c.) is applied to  layer $L-1$ which outputs and generates AVMs as $AVM = \sigma_2 (W_V^T F_A^{L-1})$, where $F_A^{L-1}$ is the extracted feature vector  at  layer $L-1$. The module applies a convolution layer to blend the $F_A$ features, then employs the sigmoid activation function. Finally, using a threshold  $\tau$ (we used $0.5$ in our study), BAM  selects the most activated regions as AVMs, indicating cancer regions.

\subsection{Loss function}

The proposed loss function $LOSS$ consists of a constraint on current year image reconstruction --- the  similarity index measure(SSIM) reconstruction loss \citep{zhao2016loss}, a layer wise constraint on the probability  of binary labels (normal and cancer) of images $y$ --- the binary cross function (BE), and a constraint on weights --- the $L2$ norm function, denoted as follows:

\begin{equation}
\label{equ:ufcnloss}
    LOSS = \underbrace{\frac{1}{N}\sum_{p\in P} 1 - SSIM(p)}_\text{Recons}  + \underbrace{\lambda_1 \sum_{l \in L}BE(y, \hat{y})}_\text{Prob} + \underbrace{\lambda_2 \sum_{w_i\in \mathbf{w}} w_{i}^2}_\text{$L_2$ Regularizer}.
\end{equation}
where 
\begin{equation}
    SSIM(p) = \frac{2\mu_i \mu_j + C_1}{\mu_i^2 + \mu_j^2 +  C_1} \cdot \frac{2\sigma_{ij} + C_2}{\sigma_i^2 + \sigma_j^2 +  C_2},
\end{equation}
where $\mu$ represents the mean of pixel intensities, $\sigma$ in this equation denotes the standard deviation of pixel intensities, $C_1$ and  $C_2$ are constant for stability. $C_1$ is given by $C_1 = (K_1T)^2$ and $C_2 = (K_2T)$ where $K_1$ and $K_2$ are constant values, and $T$ is the dynamic range of pixel intensities.

The probability distribution constraint is defined as: 

\begin{equation}
    BE(y, \hat{y}) = -( y\log(\hat{y}) + (1-y)\log(1-\hat{y})),
\end{equation}
where $y$ is the binary label for the $h(x)$ at ASG, $\hat{y}$ is prediction of the $h(x)$, and $x$ is input feature map.

\section{Baseline and Variant models}
\label{sec:baseline}
In the course of the experiments, we used the following models as baseline models and variants of the proposed model. All baseline models are supervised autoencoder-shaped models. We trained the baseline models,  proposed model, and variation models with the same training dataset and input dimension. However, the baseline models were only trained using cancer images due to their design. To have a fair comparison, all baseline models and variation models employed the same numbers of building blocks.

\subsection{U-Net}

We compared the performance of the proposed method with that of the  U-Net \citep{long2015fully}. We kept the structure of U-Net as standard U-Net and optimized the feature depth at each building block. The U-Net model has contained five building blocks.  The feature depth of each building block is indicated as $64, 128, 256, 512, 1024$. We used dice loss (Equation~\ref{equ:dice}) to optimize the U-Net gradient as follow:

\begin{equation}
\label{equ:dice}
    L_{Dice} = 1 - \frac{\sum_{n=1}^N s_n r_n + \epsilon}{\sum_{n=1}^N s_n + r_n + \epsilon} - \frac{\sum_{n=1}^N(1-s_n)(1-r_n)+\epsilon}{\sum_{n=1}^N 2-s_n-r_n + \epsilon},
\end{equation}
where $N$ is the number of images in this equation, $s$ is predicted probability, $r$ is ground truth, and $\epsilon$ is a hyperparameter to ensure the stability of the loss function. 

\subsection{Attention U-Net}

We also compared the performance of the  proposed model with that of  U-Net attention\citep{oktay2018attention}. The structure of attention U-Net reminded the same as its original and we optimized the feature depth at each building block. The feature depths of  building blocks are $64, 128, 256, 512, 1024$. 
We used dice loss (Equation~\ref{equ:dice}) to optimize the attention U-Net gradient.

\subsection{U-Net++}

Another baseline model we used in this study is U-Net++ \citep{zhou2018unet++}. The structure of the U-Net++ reminded the same. We used feature depth of $32, 64, 128, 256, 512, 64$ for building blocks. U-Net++ is an extension of the U-Net architecture for semantic image segmentation. The model structure is similar to U-Net, but with additional nested and dense skip connections.  We used dice loss (Equation~\ref{equ:dice}) to optimize the attention U-Net++ gradient.

\subsection{SegResNet}

We used SegResNet \citep{myronenko20193d} for performance comparison too. SegResNet is a deep neural network architecture designed for semantic image segmentation tasks. The model is based on the ResNet \citep{he2016deep} architecture. The SegResNet enhance the performance of ResNet for image segmentation tasks by adding a decoder network to the architecture. This decoder network is composed of several deconvolutional (or transposed convolutional) layers, which upsample the features extracted by the ResNet encoder and generate a pixel-wise segmentation mask. We used dice loss (Equation~\ref{equ:dice}) to optimize the SegResNet gradient.

\subsection{V-Net}
We also compared the proposed model with V-Net \cite{milletari2016v}. The V-Net architecture bears some resemblance to the U-Net architecture, but with some differences. Firstly, V-Net does not employ Batch Normalization, unlike U-Net. In addition, while U-Net uses element-wise summation after each successive convolutional layer, V-Net does not. In the experiment, we kept the same structure as the original V-Net. We used dice loss (Equation~\ref{equ:dice}) to optimize the V-Net gradient.

\subsection{UFCN-variants}

As mentioned in the Method 
section (section~\ref{sec:Method}), the activation function is important to obtain accurate abnormal variation maps. Therefore, to explore how the activation function impacts the proposed model, we studied two variations of the proposed model: 1) UFCN-T, and 2) UFCN-R. In the UFCN-T model, We define a new activation function, which we called it Tilu, to enhance the activated region by dropping the low-signaled neurons. The Tilu activation function can be expressed as $tiLU(x) = max(\lambda, x)$, where $\lambda$ is a small constant value as hard floor. The UFCN-R used regular ReLU activation function in the entire model. The loss function remains the same as that of the proposed method expressed in Equation~\ref{equ:ufcnloss}.

\section{Data \& Experimental Setup}

\begin{figure*}
\centering
   \includegraphics[width=1.\textwidth]{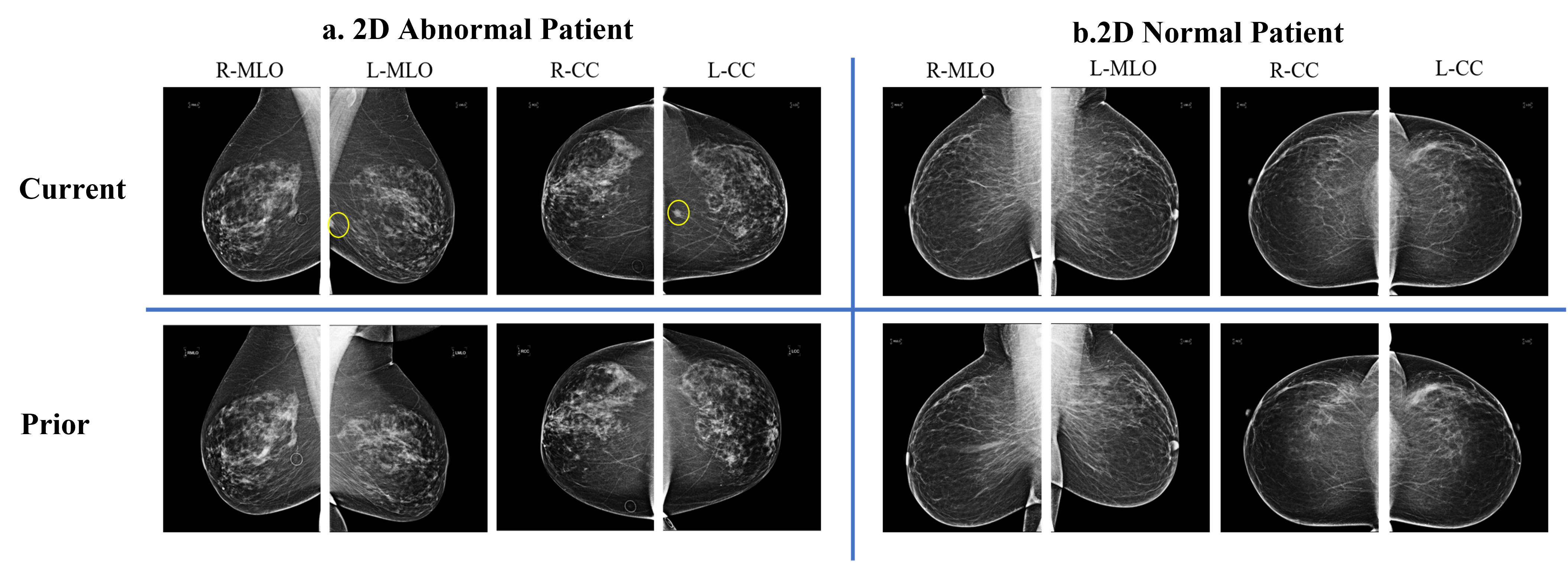}
    \caption{ Examples of current and prior mammograms of a.  an abnormal patient and  b. a normal patient from our dataset. The yellow circle indicate the abnormal location. }
    \label{fig:data}
\end{figure*}
\subsection{Experiment setup}

We used PyTorch to implement the proposed method, its variants and the baseline models. Data pre-processing was performed with High-Performance Computing (HPC) with 36 cores Xeon CPU. The proposed method was trained on Xsede with multiple 32 GB V100 GPU nodes \citep{brown2021bridges}. We explored the starting learning rate in a range from $1\mathrm{e}{-2}$ to $1\mathrm{e}{-5}$. We used a learning rate scheduler to optimize the model's learning rate. The proposed model code is available at \url{https://github.com/NabaviLab/ufcn}.

\subsection{Data}

The proposed and baseline models were trained, tested, and validated on the UConn Health Center (UCHC) dataset, which includes both current and historical mammograms. This dataset was compiled by collecting FFDMs from patients who underwent mammogram exams at UCHC between October 31, 2006, and August 23, 2021, using a Hologic machine. The UCHC Institutional Review Board approved the data collection, and the Diagnostic Imaging Informatics Department at UCHC provided support in exporting the DICOMs from the Picture Archiving and Communication Systems (PACS) at UCHC. To ensure patient privacy, identifiers were removed and replaced with a standard naming convention. The mammograms in the dataset were annotated by radiologists.

In this collection, 493 mammogram pairs (current and their corresponding prior normal mammograms) are labeled cancer, and 581 mammogram pairs are labeled normal. Data pre-processing includes normalization, re-scale, and augmentation are applied. No alignment is performed in this study. The ground truth used for evaluation is annotated by radiologists. 


To ensure the diversity and generalizability of the dataset, various types of tumors and breast densities were included. The mass type in the dataset includes round, oval, architectural distortion, irregular, and lobulated, while the microcalcification type includes amorphous, coarse, fine linear branching, pleomorphic, punctate, and round with regular shapes. All types of breast densities, including fatty, fibroglandular dense, heterogeneously dense, and extremely dense breasts, were also included. The fibroglandular dense and heterogeneously dense breast types cover a significant portion of the dataset.

\section{Results and discussion}
\label{Results}

We compared the performance of the proposed model and its variant with the baseline model explained in Section\ref{sec:baseline}, in terms of Dice score, cancer detection rate (cDR), and normal detection rate(nDR) for different cancer types (Mass, microcalcifications (Calc), Architectural Distortion (AD)). The results for different cancer types are shown in Table~\ref{tab:performance}.

We define cancer detection as binary detection when for cancer cases a cancer region is detected and for normal cases, a minimal region that below the threshold as shown in Equation~\ref{equ:TN} is detected . Additionally, we compared the performance of all the models in terms of Accuracy, Sensitivity, Precision, and F1 in detecting abnormalities. 
 True Positive ($TP$) and True Negative ($TN$) used in computing the aforementioned metrics are defined  in Equations~\ref{equ:TP} and~\ref{equ:TN}, respectively.

\begin{equation}
\label{equ:TP}
    TP =\begin{cases}
    1, & \text{if $dice(I)>0.01$}\\
    0, & \text{otherwise}
  \end{cases},
\end{equation}
where $I$ is the input test mammogram image.

\begin{equation}
\label{equ:TN}
    TN =\begin{cases}
    1, & \text{if $\sum_{i,j\in w, h} I(i,j) <0.01$}\\
    0, & \text{otherwise}
  \end{cases},
\end{equation}
where $w, h$ are the width and height of the input image. 

\subsection{Overall results}

Table~\ref{tab:performance} highlights the superiority of the proposed model, UFCN, in terms of cancer detection. This model achieves the best cancer detection rate for both masses and microcalcifications, as well as the best Dice score for masses. Furthermore, the proposed UFCN also performs well in architectural distortion cancer type, achieving the second-best performance in terms of the Dice score and the third-best in terms of cancer detection rate. When compared to its variant, UFCN-R, which uses the ReLU activation function, the proposed UFCN outperforms in all types of cancers. The inferior performance of UFCN-R may be caused by the dying ReLU, in which ASG suppresses the majority of nodes to zero, resulting in a decrease in the activated regions and shrinkage of the abnormal variation map. In contrast, the  use of the "soft floor" SiLU activation function in UFCN prevents dead neurons and provides the necessary stability to activate the attention regions while simultaneously suppressing the irrelevant background of cancer images and normal images. The results demonstrate the superior performance of UFCN in cancer detection, even though the model is trained in an unsupervised fashion.

For the architectural distortion cancer type, UFCN-T shows superior performance in both Dice score and cancer detection rate. Although the TiLU activation function uses a hard floor like ReLU, it employs a slightly tighter bound as a hard floor to avoid the dying ReLU caused by the properties of zero, unlike the ReLU activation function that uses $0$ for the hard floor. In other words, UFCN-T maximizes the variation between the current year image and the prior image, which is also observable by its superior performance in detecting microcalcifications and comparable performance in detecting masses. 

The proposed model, UFCN, outperforms all baseline models including U-Net, U-Net attention, U-Net++, SegResNet, and V-Net in almost all the metrics. Among all baseline models, U-Net++ showed the best performance in terms of cancer detection rate and Dice score in architectural distortion and mass cancer types, while U-Net showed a decent Dice score and cancer detection rate in Cals among all baseline models. However, both models showed relatively low performance in normal tissue classification, as indicated by their low nDR. In contrast,  UFCN achieved the best performance in terms of cancer detection rate and Dice score for mass and the second-best for calcifications, as well as the highest nDR among all the models. These findings demonstrate the superior performance of our proposed UFCN, which is trained in an unsupervised fashion. Specifically, UFCN-T showed the best performance overall, with a Dice score of 0.66 and a cDR of $0.86$, followed by UFCN with a Dice score of $0.69$ and a cDR of $0.91$.  These findings demonstrate the superior performance of our proposed UFCN, which is trained in an unsupervised fashion.


\begin{table}[]
    \centering
   \caption{Models' performance on  localizing abnormalities and detection rate normal and cancer}

    \begin{tabular}{|p{85pt}|p{30pt}|p{30pt}|p{30pt}|p{30pt}|p{30pt}|p{30pt}|p{30pt}|}
    \cline{2-8}
        \multicolumn{1}{c|}{}&\multicolumn{2}{c|}{AD}&\multicolumn{2}{c|}{Mass}&\multicolumn{2}{c|}{Cals}& \multicolumn{1}{c|}{Normal}\\
       \hline
         Model & Dice & cDR & Dice & cDR   & Dice & cDR & nDR  \\\hline 
         U-Net& 0.47&0.71&0.64&0.87  &0.44&0.71&0.09\\\hline
         U-Net attention&0.48 & 0.79& 0.66&0.87&0.40 &0.68&0.09\\\hline
          U-Net++& 0.60 &\textbf{0.86}&0.68&0.89  &0.38&0.58&0.08\\\hline
           SegResNet& 0.58&\textbf{0.86}&0.63&0.85  &0.39&0.61&0.21\\\hline
            V-Net& 0.46&0.64&0.58&0.75  &0.29&0.45&0.08\\\hline
         
         \hline\hline
         UFCN-T&\textbf{0.66} & \textbf{0.86}&0.61 & 0.76&\textbf{0.59} & 0.61 &0.17\\\hline
         UFCN-R& 0.35& 0.64& 0.58& 0.73& 0.36& 0.55&0.58\\\hline
         UFCN& 0.60& 0.79& \textbf{0.69} & \textbf{0.91}&0.57 & \textbf{0.74}&\textbf{0.73}\\\hline

    \end{tabular}
   
    \label{tab:performance}
\end{table}

In addition to evaluating cancer detection rates and Dice scores, we also assessed the model's performance in detecting normal and cancer cases in terms of Accuracy, Sensitivity, Precision, and F1 scores (see Table~\ref{tab:accPerformance}). 
Notably, all the baseline models do not perform well when applied to normal cases. The baseline models originally proposed to identify abnormal tissue areas in cancer images, and it often relies on pre-classified cancer data. This limitation causes the baseline models to fail to distinguish between cancer and normal cases. As a result, the U-Net model shows poor performance in terms of Accuracy ($0.41$), Sensitivity ($0.43$), and F1 score ($0.56$) but a better Precision score ($0.80$). The SegResNet showed slightly better performance in Accuracy ($0.47$) and Sensitivity ($0.46$) compare with U-Net. The V-Net shows least performance in terms of all evaluation metrices.

\begin{table}[]
    \centering
   \caption{Models' cancer and normal detection performance}
\begin{tabular}{|p{85pt}|p{45pt}|p{45pt}|p{45pt}|p{45pt}|}
\hline
Model  & Accuracy & Sensitivity & Precision & F1 \\ \hline
U-Net  &     0.41     & 0.43            &0.80           & 0.56   \\ \hline
U-Net attention & 0.42& 0.43&0.80 & 0.56\\\hline
U-Net++ &0.40 & 0.42& 0.79& 0.55\\\hline
SegResNet &0.47 & 0.46& 0.78& 0.58\\\hline
V-Net & 0.34& 0.37& 0.64&0.47 \\\hline \hline

UFCN-T &   0.43       &   0.43          &0.73           &  0.54  \\ \hline
UFCN-R &   0.62       &      0.57       &0.67           &   0.62 \\ \hline
UFCN &  \textbf{0.78}        &  \textbf{0.72}           &  \textbf{0.84}         & \textbf{0.78} \\ \hline
\end{tabular}
\label{tab:accPerformance}
\end{table}

Although UFCN-T shows a higher detection rate for architectural distortion and microcalcification, its normal detection rate is lower, and its accuracy ($0.43$), sensitivity ($0.43$), and F1 score ($0.54$) are comparable with those of U-Net. Similar to U-Net, the precision score of UFCN-T yields the third-best result. UFCN-R shows a better normal detection rate compared to U-Net and UFCN-T. However, the trade-off to increasing the normal detection rate in UFCN-R is a lower detection rate for cancer cases. Hence, its accuracy ($0.62$), sensitivity ($0.57$), and F1 score ($0.62$) are the second-best results. However, its precision is the lowest compared to those of the other models. The proposed UFCN shows the best performance in terms of all the evaluation metrics compared to the other models. The UFCN achieves the best normal detection rate ($0.73$) while still maintaining a better performance for cancer detection. As Table~\ref{tab:accPerformance} demonstrates,  UFCN shows the best accuracy ($0.78$), sensitivity ($0.72$), precision ($0.84$), and F1 score ($78$).

\subsection{Cancerous case results}

\begin{figure*}
\centering
   \includegraphics[width=1\textwidth]{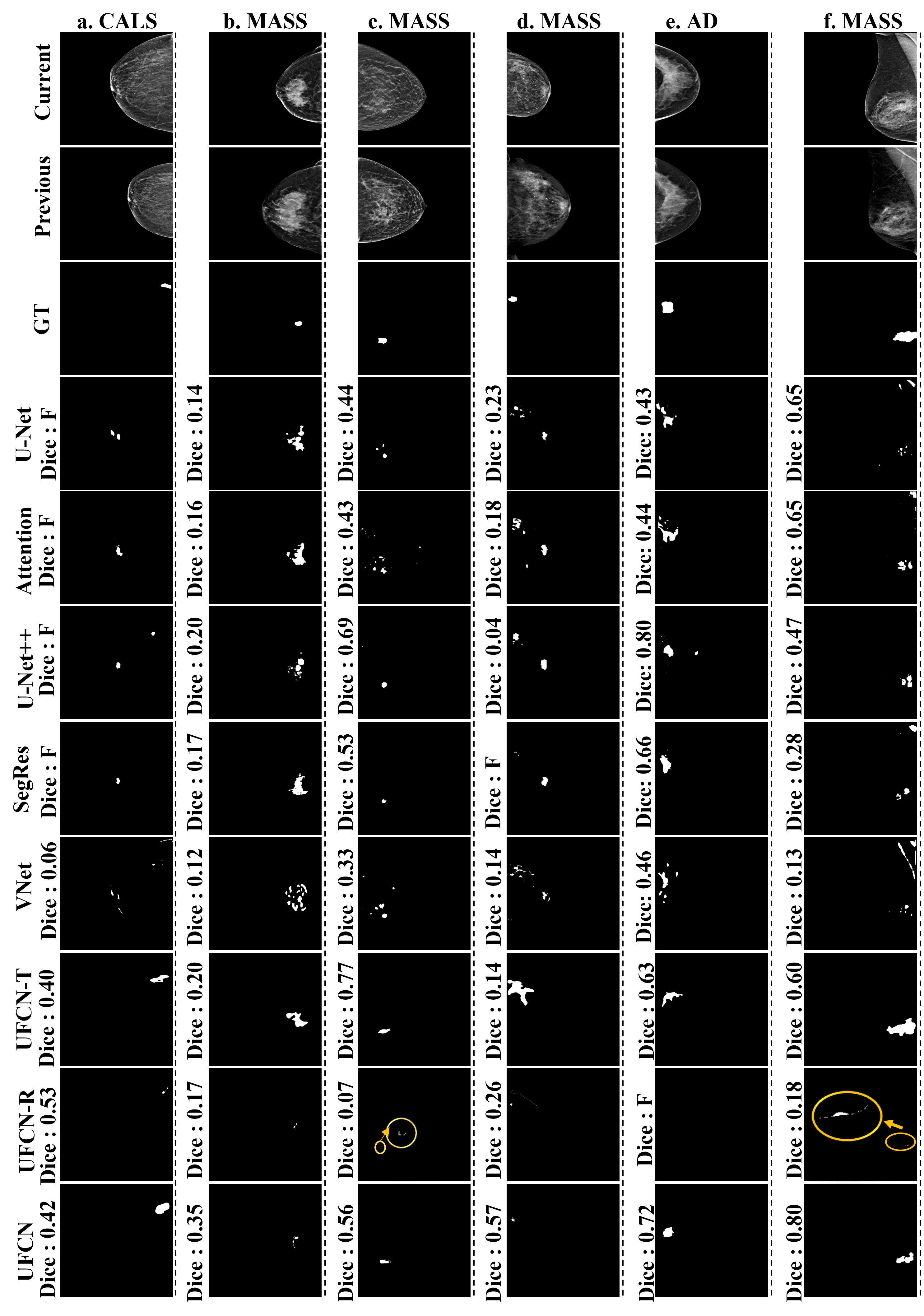}
    \caption{Outputs of the baseline and proposed models. The first row shows current year, second row shows prior year, and third row shows ground truth (GT) images. The yellow circle is enlarged area for better visualization. F indicate failed to localize abnormal tissue.}
    \label{fig:abnormal1}
\end{figure*}

We examined the  AVM outputs of the propose model and its variant and also the  segmentation outputs of the baseline models. 
As shown in Figure~\ref{fig:abnormal1}, UFCN's prediction is very close to the ground truth annotations. Especially for architectural distortion (~\ref{fig:abnormal1}.e.) and microcalcification (~\ref{fig:abnormal1}.a.), the UFCN can generates more precise abnormal tissue maps (Figure~\ref{fig:abnormal1} a \& e). As can be seen in Figure~\ref{fig:abnormal1}.a, U-Net, attention U-Net, U-Net++ and SegResNet models fail to detect microcalcification. UFCN-R misses the detection for architectural distortion. As mentioned before, the hard floor of ReLU causes the active region to shrink as shown in the Figure. UFCN-T generates larger abnormal tissue areas compared with UFCN and UFCN-R. 



\begin{figure*}
\centering
   \includegraphics[width=1.\textwidth]{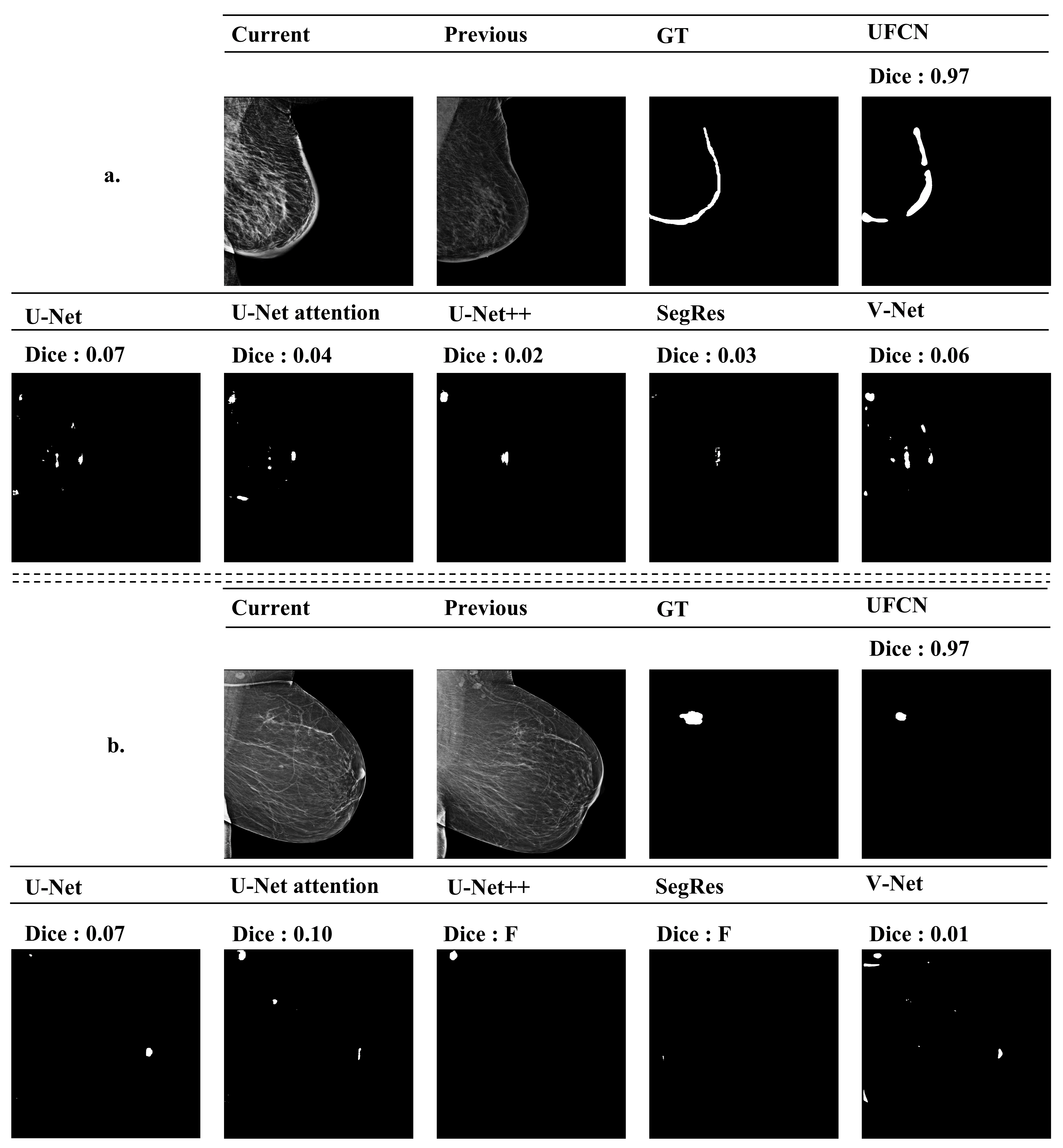}
    \caption{A demo for special cases. a. The abnormal tissue area is located at breast skin. b. There is a normal high intensity region. }
    \label{fig:skin}
\end{figure*}

What stands out in our results is the skin tumor case as shown in Figure~\ref{fig:skin}.a. This case was only in our test data set. As the figure shows, the U-Net model misses the skin area and wrongly detects a few other regions as tumor locations. On the other hand, the proposed UFCN model is able to precisely generate the abnormal region, which indicates that the ASG module  in the proposed model was not activated in the breast tissue area, but instead it was activated in the area where the abnormal changes were exist. This demonstrates the effectiveness of the ASG mechanism in accurately localizing the abnormal tissue region in medical images. Furthermore, this finding highlights the importance of developing models that are specifically designed to handle diverse abnormalities in medical images. The proposed UFCN model demonstrates superior performance in detecting abnormal tissue areas, including those that may not be related to the specific medical condition under consideration, such as in the case of skin tumors.

In Figure~\ref{fig:skin}.b., we visualized the output maps generated by all baseline models and our proposed UFCN on a mass case. As can be observed 
in Figure~\ref{fig:skin}.b., U-Net mistakenly identified a bright round area as an abnormal tissue region, leading to false negative detection. This limitation is a common issue with U-Net, where non-cancerous bright round areas are frequently misidentified as abnormal.  U-Net++ and SegResNet also failed to detect cancer in this particular case. In contrast, our proposed model, UFCN, effectively distinguishes between normal and abnormal bright areas by comparing current and previous images. Consequently, the abnormal variation map (AVM) generated by UFCN shows more precise and accurate cancer detection.


\subsection{Non-cancerous results}

\begin{figure*}
\centering
   \includegraphics[width=1\textwidth]{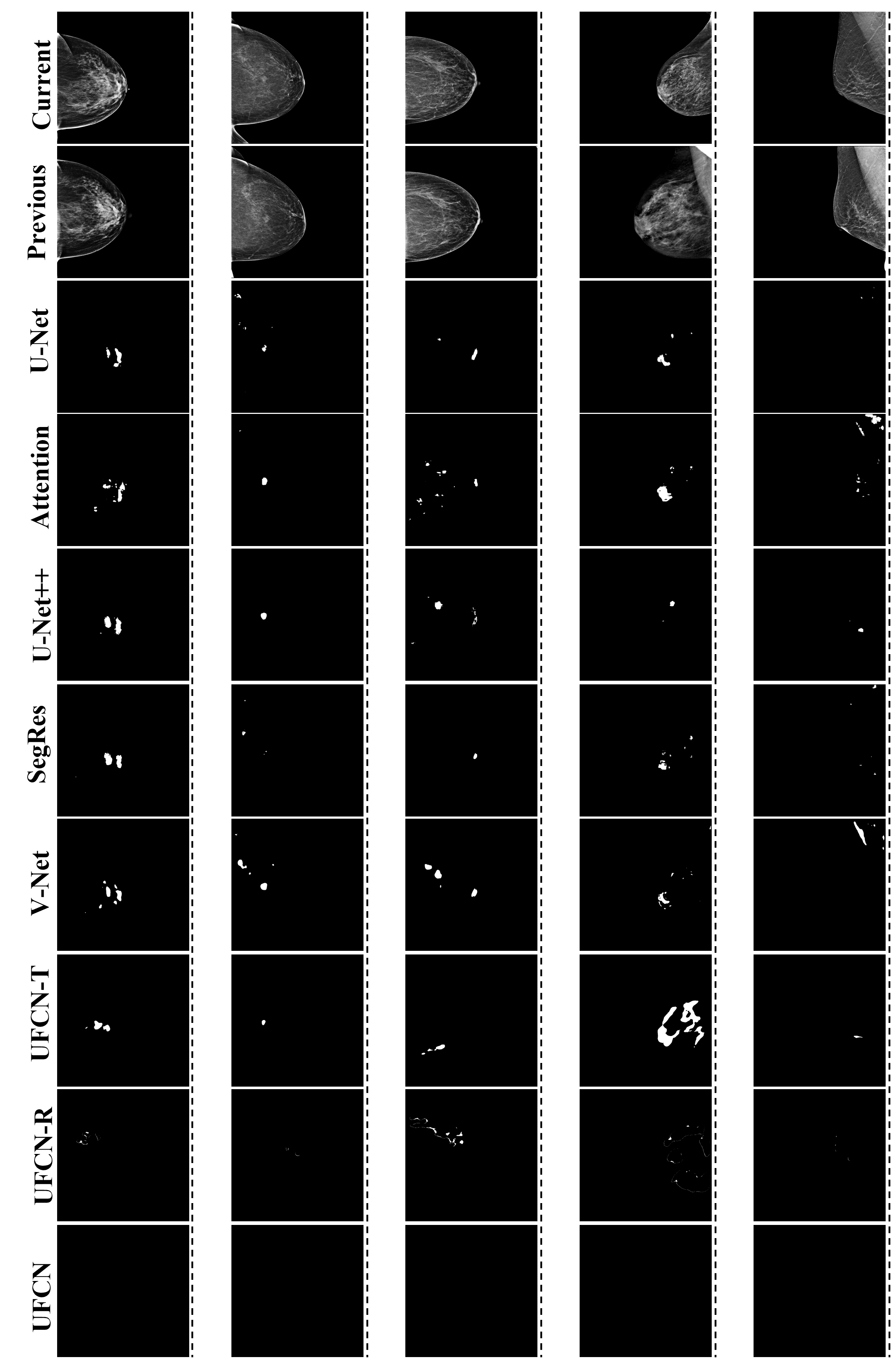}
    \caption{The models' output of baseline model and proposed models. The first column is current year image, second column is prior year. The black output indicate no any suspicious tissue detected.}
    \label{fig:normal}
\end{figure*}

The Figure~\ref{fig:normal} illustrates the prediction of all the models when inputs are  normal cases. All baseline models showed detected tissues, and the wrongly detected tissue locations are similar in all models. Although UFCN-R showed minor wrongly detected tissue areas compared with U-Net and UFCN-T, it still performed relatively well. However, the proposed model UFCN showed no detection of abnormal tissues  in these normal cases, indicating its high accuracy in distinguishing between normal and abnormal cases. 

 In summary of our study, we demonstrate that by learning the differences between the current and prior images we can achieve unsupervised cancer area localization. Labeling in medical image studies is expensive and prone to errors, making unsupervised learning an ideal approach. Our results show that with prior images, the proposed model can achieve results as good as those of a supervised model. Additionally, the proposed model outperformed the supervised model in detecting complex tumors that the latter was unable to detect.

Overall, our study highlights the potential of unsupervised learning for medical image analysis, particularly for tasks such as cancer area localization. By leveraging longitudinal data and advanced machine learning techniques, we can reduce the need for costly and time-consuming manual labeling while still achieving high levels of accuracy and sensitivity in cancer detection. These findings have important implications for improving the efficiency and effectiveness of cancer screening and diagnosis, ultimately leading to better patient outcomes.     

\section{Conclusion}
\label{others}

Breast cancer remains a leading cause of death for women worldwide. Early detection and accurate diagnosis of breast abnormalities are crucial for improving patient outcomes. However, traditional screening methods such as mammography have limitations in terms of accuracy and sensitivity, leading to missed or misdiagnosed cases. One of the main challenges in detecting cancer is the lack of large annotated datasets to train advanced segmentation models.

To address this issue, we developed a novel unsupervised feature correlation network to predict breast abnormal variation maps using 2D mammograms. Our proposed model takes advantage of the reconstruction process of the current year and prior year images to extract tissue from different areas without a need for ground truth. By analyzing the differences between the two images, our model can identify abnormal variations that may indicate the presence of cancer.

The model is embedded with a novel feature correlation module, an attention suppression gate, and a breast abnormality module, all of which work together to improve the accuracy of the prediction. The feature correlation module allows the model to identify patterns and relationships between different features, while the attention suppression gate helps to filter out irrelevant information. The breast abnormality module then uses this information to classify the input as normal or cancerous.

Notably, our proposed model not only provides breast abnormal variation maps but is also able to distinguish between normal and cancer inputs, making it more advanced compared to the state-of-the-art segmentation models. The state-of-the-art segmentation models need already classified cancer images, which requires applying a classification method first, then, using the segmentation method. The results of our study show that the proposed model outperforms or performs as well as the supervised state-of-the-art segmentation models not only in localizing abnormal regions but also in recognizing normal tissues.

\section*{Acknowledgments}
This work is supported by a grant from the University of Connecticut Research Excellence Program, PIs: Nabavi and Yang. This work used the Bridges-2 system at the Pittsburgh Supercomputing Center (PSC) as part of the Extreme Science and Engineering Discovery Environment (XSEDE) which is supported by the National Science Foundation, under project number CIS220028, PI: Nabavi.

\bibliographystyle{model2-names}\biboptions{authoryear}
\bibliography{elsarticle-template-num}

\end{document}